\documentclass[prl,aps,twocolumn,groupaddress]{revtex4-1}
\usepackage{latexsym}
\usepackage{hyperref}
\usepackage{url}
\usepackage{graphicx}
\usepackage{verbatim}
\usepackage{multirow}
\usepackage{amsmath}
\newcommand{\beq}{\begin{eqnarray}}
\newcommand{\eeq}{\end{eqnarray}}
\usepackage{mathrsfs}
\usepackage{float}
\usepackage[usenames, dvipsnames]{color}
\usepackage{mathtools}
\usepackage{slashed}
\usepackage{physics}	
\usepackage{graphicx}   
\usepackage{epstopdf}
\usepackage{subfigure}  
\usepackage{hyperref}   
\usepackage{bbold}
\usepackage{wasysym}
\usepackage{feynmp}
\usepackage[symbol]{footmisc}
\def \bs{\textbf}
\usepackage[latin1]{inputenc}
\usepackage{tikz}
\usetikzlibrary{decorations.pathmorphing}
\usetikzlibrary{shapes.misc}
\tikzset{cross/.style={cross out, draw=black, minimum size=8*(#1-\pgflinewidth), inner sep=0pt, outer sep=0pt},
cross/.default={1pt}}
\begin{document}

\title{Glass-induced enhancement of superconducting $T_c$: Pairing via dissipative mediators}
\author{Chandan Setty}
\thanks{email for correspondence: csetty@ufl.edu}
\affiliation{Department of Physics, University of Florida, Gainesville, Florida, USA}

\begin{abstract}
With substantial evidence of glassy behavior in the phase diagram of high $T_c$ superconductors and its co-existence with superconductivity, we attempt to answer the question: what are the properties of a superconducting state where the force driving cooper pairing becomes dissipative? We find that when the bosonic mediator is local, dissipation acts to reduce the superconducting critical temperature ($T_c$). On the other hand, contrary to na\"{i}ve expectations,  $T_c$ behaves non-monotonically with dissipation for a non-local mediator -- weakly dissipative bosons at different energy scales act coherently to give rise to an increase in $T_c$  and eventually destroy superconductivity when the dissipation exceeds a critical value. The critical value occurs when dissipative effects become comparable to the energy scale associated with the spatial stiffness of the mediator, at which point, $T_c$ acquires a maximum. 
We outline consequences of our results to recent proton irradiation experiments (M. Leroux et al.,~\cite{Welp2018}) on the cuprate superconductor La$_{2-x}$Ba$_x$CuO$_4$ (LBCO) which observe a disorder induced increase in $T_c$ even when the transition temperature of the proximate charge density wave (CDW) seems to be unaffected by irradiation. 
Our mechanism is a novel way to raise $T_c$ that does not require a `tug-of-war' -like scenario between two competing phases.  
\end{abstract}

\maketitle
\section{Introduction} 
In $s$-wave superconductors (SCs) where the quasiparticle excitation spectrum is fully gapped and has a constant sign of the pairing form factor across the Fermi surface, Anderson's magic theorem keeps the critical temperature ($T_c$) robust to weak, non-magnetic impurities. In higher angular momentum SCs ($p$-,$d$-wave etc) or SCs where the sign of the gap changes across parts of the Fermi surface (such as $s_{\pm}$ pnictide SCs), $T_c$ is drastically suppressed with the addition of  impurities$-$magnetic or otherwise~\cite{Zhu2006}. These effects hold in the independent disorder limit and in the absence of electron correlations. \par
At a collective level when electron correlations are taken into account, randomness can yield several interesting phases of matter~\cite{Ramakrishnan1985}. Amongst these is the spin glass (SG) phase widely observed in the phase diagram of many strongly correlated systems like high $T_c$ SCs~\cite{Aharony1995, Yamada2000, Ferretti2001, Imai2001, Julien2003, Birgeneau1990, Yamada2008, Hirota2018, Lin1999, Grafe2012, Julien2013, Curro2013, Mesot2013, Petrovic2015-Co, Petrovic2015-Ni, Buechner2012, Paulose2010}. The SG phase exhibits a  remarkable phenomenology~\cite{Mydosh2015} -- a transition into the SG defined by a broad cusp in the specific heat, a split in the DC magnetization at the SG transition depending on whether the SG phase is field cooled (FC) or zero field cooled (ZFC), linear temperature dependence of the AC susceptibility peak, and aging. Theoretically, SGs are described by an order parameter where the  spin average on each site is non-vanishing but goes to zero when averaged over the lattice~\cite{Ye1995}.  Important to the discussions that follow, spin correlators at the SG critical point follow a power law of the form~\cite{Moore1985, Huse1993, Ye1995, Phillips1999}
\beq
D(\tau) \equiv \bigg[\langle S_{i\mu}(\tau)S_{i\mu}(0)\rangle \bigg] \sim \frac{1}{\tau^2},
\eeq 
which in frequency space reads $D(\omega)\sim |\omega|$. Here, $S_{i\mu}$ is the $\mu-$th component of the spin at site $i$, and the angular and square brackets denote thermal and site averages respectively. The linear frequency dependence of the spin correlators indicates that dissipative dynamics is a necessary -- albeit not sufficient -- ingredient of SGs. \par
In this work, we explore the robustness of $T_c$ and properties of a superconducting state where the dynamics of the pairing mediator is rendered dissipative due to \textit{collective} disorder (in the aforementioned sense). To this end, we add to the Lagrangian describing the mediator a dissipative term~\cite{Leggett1983, Halperin1987, Zimanyi1988, Zimanyi1997} $\Delta L =  \sum_{\bs k, \omega_n} \eta|\omega_n||\Psi(\bs k, \omega_n)|^2$. Here $\bs k$ and $\omega_n$ are the momenta and Matsubara frequencies, $\eta$ is a measure of dissipation, and $\Psi(\bs k, \omega_n)$ is the bosonic field. We find that dissipative effects generally act to suppress $T_c$ when the mediator is local. This occurs because dissipation has the effect of reducing the attractive interaction mediating cooper pairs. However, contrary to na\"{i}ve expectations,  $T_c$ behaves non-monotonically with dissipation for a non-local mediator. In this scenario, weakly dissipative bosons at different energy scales act coherently to give rise to an increase in $T_c$  and eventually destroy superconductivity when the dissipation exceeds a critical value. The critical value occurs when the dissipation parameter, $\eta$, becomes comparable to the energy scale associated with the velocity of the mediating bosons (or the spatial stiffness); at this crossover, $T_c$ acquires a maximum value. We also study the effects of dissipative mediator on the  ratio $\frac{2\Delta(0)}{T_c}$ and the heat capacity jump at the superconducting transition and find departures from values predicted by BCS theory.   \par
\section{Experimental basis} We now make our case for a dissipative or `glassy' mediator from experiments on a variety of high $T_c$ SCs. The SG phase has been observed extensively in the under-doped and regions proximate to superconductivity in the phase diagrams of both the cuprate~\cite{Aharony1995, Yamada2000, Ferretti2001, Imai2001, Julien2003, Birgeneau1990, Yamada2008, Hirota2018, Lin1999, Grafe2012, Julien2013} and iron based superconductors~\cite{Curro2013, Mesot2013, Petrovic2015-Co, Petrovic2015-Ni, Buechner2012, Paulose2010}. Existing evidence is also spread over several techniques such as DC magnetization~\cite{Aharony1995, Buechner2012, Paulose2010, Petrovic2015-Co, Petrovic2015-Ni, Yamada2000}, NMR/NQR~\cite{Curro2013, Hirota2018, Imai2001, Grafe2012, Julien2013, Lin1999}, $\mu$SR~\cite{Birgeneau1990, Yamada2008} and neutron scattering~\cite{Birgeneau1990, Mesot2013}. Given the strong evidence of a SG phase and its proximity to the superconducting dome in high $T_c$ SCs, it is already reasonable to consider its effect on the pairing problem. Additionally, there is ample experimental evidence lending credence to a dissipative character of fluctuations that mediate Cooper pairing. First, disorder causes the $d$-electron spins (Cu spins in the cuprates and Fe spins in the iron superconductors) to exhibit glassy behavior and not the dopant spins~\cite{Julien2003, Aharony1995, Curro2013, Buechner2012} (although in certain iron based systems, it is the dopant spins become glassy~\cite{Qiu2015}). Second, SG and SC phases actually co-exist in a variety of high $T_c$ SCs~\cite{Lin1999, Curro2013, Buechner2012}. This indicates a strong inter-mixing of properties of the two phases, similar to what is expected in the context of other mean-field orders (such as density waves) acquiring a glassy behavior~\cite{Senthil2015}. Third, neutron scattering and NMR/NQR measurements in the cuprate SCs La$_{2-x}$Sr$_x$CuO$_2$ (LSCO) and La$_{2-x}$Ba$_x$CuO$_4$ (LBCO) have found a direct `slowing' of spin fluctuations in the vicinity of glassy orders~\cite{Imai2001, Mesot2013, Birgeneau1990, Julien2013}. Finally, early theoretical predictions on doping La$_2$CuO$_4$ clearly point to a frustration induced glassy behavior of the Cu $d$-orbital spins in the phase diagram~\cite{Stanley1988}. \par
Hence, the notion of a dissipative pairing mediator in high temperature superconductors has firm foundations in both experiment and theory. As will be argued later in this paper, non-local dissipative mediators can help throw light on recent proton irradiation experiments~\cite{Welp2018} on LBCO which observe a disorder induced increase in $T_c$ \textit{even when the transition temperature of the proximate charge density wave (CDW) is unaffected by the presence of radiation disorder}. The mechanism we propose in this paper forms an alternative way to raise $T_c$ of a superconductor that does not require a `tug-of-war' -like scenario between two competing phases.   
\par
\section{Model and gap equation} We begin by writing the conjectured model for the bosonic propagator. The total action consists of a free part $S_0[\Psi, \Psi^*] $ and a dissipative part $ S_{dis}[\Psi, \Psi^*]$ defined by
\beq \nonumber
S[\Psi, \Psi^*] &=& S_0[\Psi, \Psi^*] + S_{dis}[\Psi, \Psi^*] \\ \nonumber
S_0[\Psi, \Psi^*] &=& \int d^d \bs r d\tau \bigg[ \kappa |\nabla \Psi(\bs r, \tau)|^2 + |\partial_{\tau} \Psi(\bs r, \tau)|^2 \\\nonumber
&& + M^2 |\Psi(\bs r, \tau)|^2  \bigg],\\ \nonumber
\eeq
\begin{figure}[h!]
\includegraphics[width=1.7in,height=1.26in]{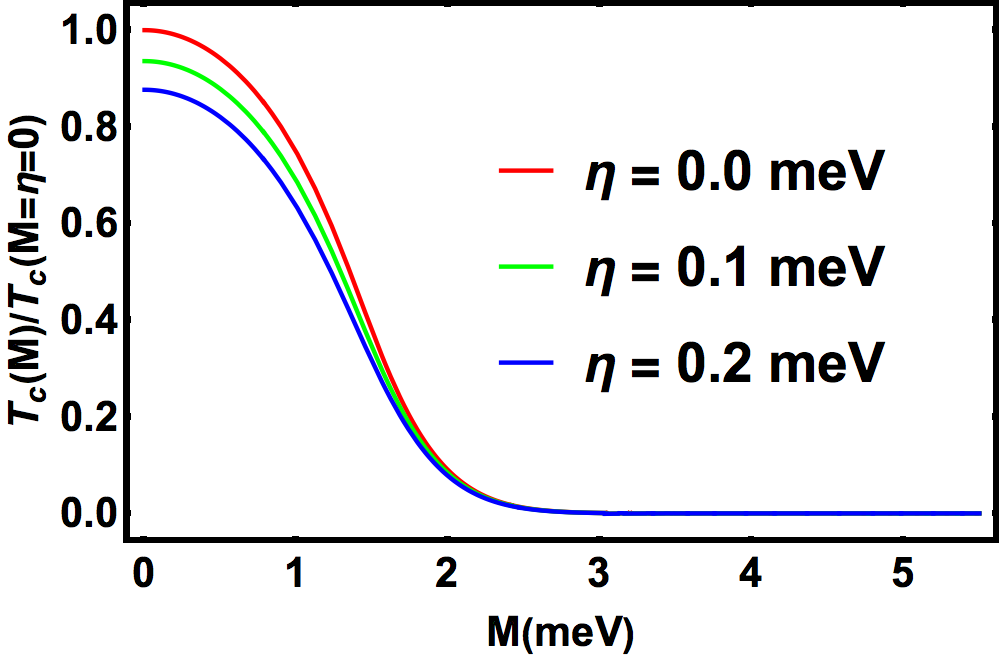}\hfill%
\includegraphics[width=1.7in,height=1.25in]{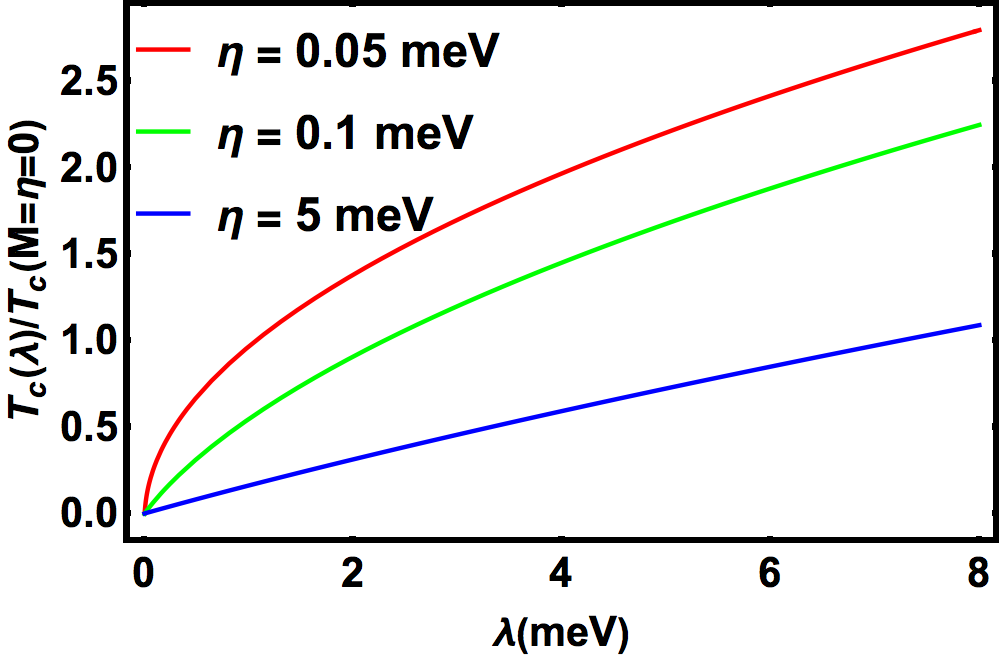}
\caption{Effect of a local dissipative mediator: (Left) Superconducting critical temperature $T_c$ (normalized to its value at $M=\eta=0$) as a function of the mass parameter $M$. The coupling constant $\lambda$ is chosen to be equal to 1 meV. (Right) Same quantity now plotted as a function of the coupling constant $\lambda$ for $M=0.1$ meV. A crossover from $T_c\sim \sqrt{\lambda}$ to  $T_c\sim \lambda$ occurs as a function of $\eta$. 
} \label{TcVsMandLambda}
\end{figure}
where $\kappa$ is the spatial stiffness or energy scale associated with the boson velocity, and the squared mass, $M^2$, is proportional to the inverse correlation length. As outlined in the introduction, we take the dissipative term to be form $S_{dis}[\Psi, \Psi^*] = \sum_{\bs k, \omega_n} (2\eta~|\omega_n|)|\Psi(\bs k, \omega_n)|^2 $ in Fourier space with the various quantities defined previously.  With this total action, the bosonic propagator, $D(\bs q, i \omega_{n} - i \omega_m)$, takes the form 
    \beq \nonumber
    D(\bs q, i \omega_{n}) = \frac{\alpha}{\kappa q^2 + \omega_n^2 + 2 \eta |\omega_n| + M^2 }.
       \eeq
Here $q = |\bs q|$ and  $\alpha$ is a constant with dimensions of energy that can be absorbed into an effective coupling constant (similar to spin fluctuations; see for example~\cite{Ketterson2008}). \par
\begin{figure}[h!]
\includegraphics[width=1.7in,height=1.25in]{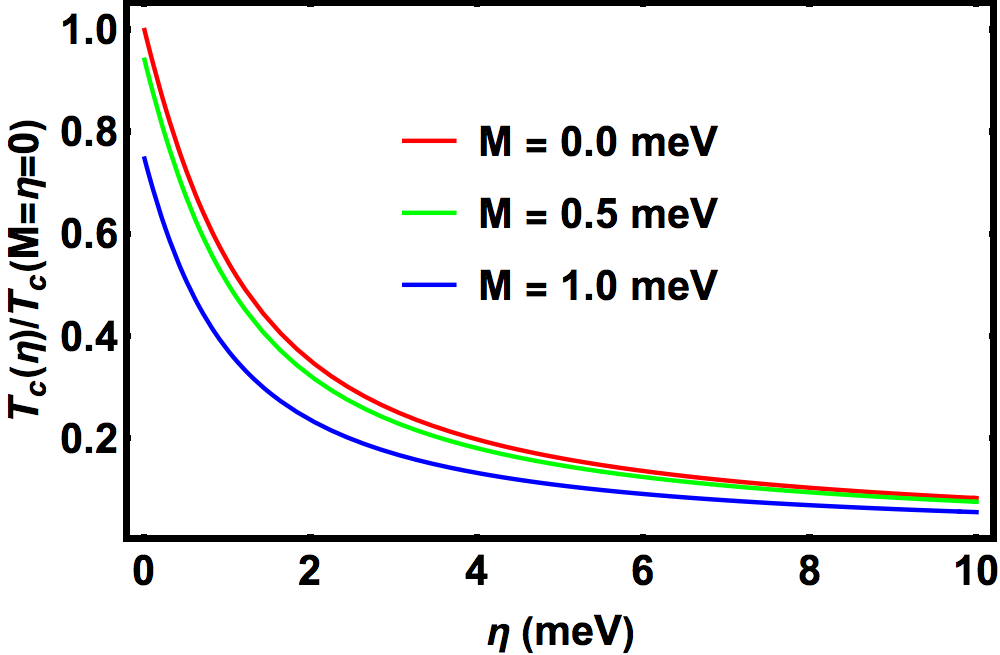}\hfill%
\includegraphics[width=1.7in,height=1.25in]{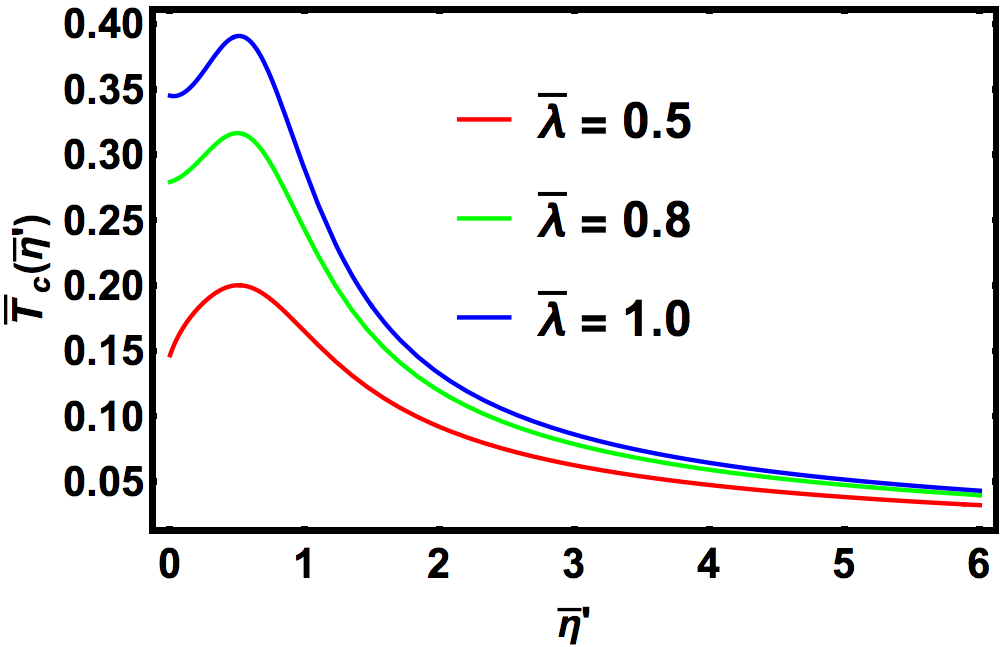}
\caption{(Left) Superconducting critical temperature $T_c$ (normalized to its value at $M=\eta=0$) as a function of the dissipation parameter $\eta$ for different masses $M$ when the mediator is local. The coupling constant $\lambda$ is chosen to be equal to 1 meV. (Right) The case when the mediator is non-local for $M=0$: plot of the dimensionless $\bar{T}_c = T_c/\kappa$ as a function of $\bar{\eta}' = \eta'/\kappa$ for different dimensionless coupling strengths $\bar{\lambda} = \lambda/\kappa^2$. The peak in $T_c$ is set by $\kappa$, the energy scale associated with bosonic velocity.
} \label{Tc-Nonmonotonic}
\end{figure}

\textit{$T_c$, local case ($\kappa=0$)}: We choose a quadratic electron dispersion $\xi_{\bs k}$ and gap function denoted by $\Delta(i\omega_n, \bs k)$. Substituting $D(\bs q, i \omega_{n} - i \omega_m)$ into the gap equation, 
\begin{align}
\Delta(i\omega_n, \bs k) = \frac{|g|^2}{\beta V} \sum_{\bs q, \omega_m} \frac{D(\bs q, i \omega_n  - i \omega_m) \Delta(i\omega_m, \bs k + \bs q)}{ \omega_m^2 + \xi_{\bs k + \bs q}^2 + \Delta(i\omega_m, \bs k + \bs q)^2},
\label{GapEquation}
\end{align}
and assuming an isotropic, frequency independent $s$-wave gap (defined by the $\bs k =0$ value and denoted by $\Delta$ henceforth), the equation determining $T_c$ (setting $\Delta =0$) reduces to $1 = \pi \lambda T \sum_{\omega_m < \Lambda} \frac{1}{|\omega_m| \left( \omega_m^2 + 2 \eta |\omega_m| + M^2\right)}$ where $T$ is set to $T_c$. Here $\beta$ is the inverse temperature, $g$ is the interaction strength, $\lambda \equiv N(0)|g|^2 \alpha$ is the coupling constant, $\Lambda$ is the high energy cut-off and $N(0)$ is the density of states at the Fermi energy. The sum over $\omega_m$ can be performed exactly to yield the equation for $T_c$ as ($c.c$ is complex conjugate)
 \beq \nonumber
 1 &=& \frac{\lambda  \left( \eta - i \bar{M}\right)^{-1}}{2 i \bar{M}} \bigg[ \psi\left(\frac{1}{2} + \frac{\eta}{2 \pi T_c} - i \frac{\bar M}{ 2\pi T_c} \right) - \psi\left(\frac{1}{2}\right) \bigg] \\
 && +~~~~ c.c ,
 \eeq
 where $\bar{M} \equiv \sqrt{M^2 -\eta^2}$ and $\psi(x)$ denotes the digamma function.  The solutions for $T_c$ as a function of the parameters $M$ and $\eta$ are shown in the left panels of Figs~\ref{TcVsMandLambda} and \ref{Tc-Nonmonotonic}. The reduction in $T_c$ as a function of the mass and dissipation can be intuitively understood by taking the limit of $M\gg T, \eta$ and $\eta\gg T, M$ respectively. In these limits, $\eta$ and $M^2$ can be factored out of the Matsubara sum which, in effect, reduces the coupling constant $\lambda$ and hence suppresses $T_c$.\par
 \textit{$T_c$, non-local case ($\kappa\neq0$)}: We can make similar assumptions on the superconducting gap for the $\kappa\neq 0$ case. To maintain analytical tractability and focus on the effect of dissipation parameter $\eta$, we will later set the mass (now renormalized by the chemical potential; we use the same symbol for ease of notation) to zero.  We can now substitute the bosonic propagator with $\kappa\neq0$ back into the gap equation Eq~\ref{GapEquation}. 
 The resulting energy integral can be solved exactly by the method of residues and takes the form $\int_{-\infty}^{\infty} \frac{d\xi}{(\xi^2 + r^2)( \kappa \xi + s )} = \frac{\pi s}{(\kappa^2 r^2 + s^2) r}$, where $r^2 = \omega_m^2 + \Delta^2$ and $s = \omega_m^2 + \eta' |\omega_m| + M^2$. Performing the remaining Matsubara sum we obtain the equation for $T_c$ as
 \beq \nonumber
 1 &=& - \lambda \Bigg[ \frac{\psi\left(\frac{1}{2} + \frac{\eta' -  i \kappa}{2 \pi T_c}\right)}{2\left(\eta' - i \kappa \right)^2} +  \frac{\psi\left(\frac{1}{2} + \frac{\eta' + i \kappa}{2 \pi T_c}\right)}{2\left(\eta' + i \kappa \right)^2} \\
 && + \frac{\kappa^2 - \eta'^2}{(\kappa^2 + \eta'^2)^2} \psi\left( \frac{1}{2} \right) - \frac{\pi^2 \eta'}{4 \pi T_c (\eta'^2 + \kappa^2)}\Bigg]
 \label{Non-Local-GapEqn}
 \eeq
where $\eta' \equiv 2 \eta$. The solution for $\bar{T}_c = T_c/\kappa$ is plotted in the right panel of Fig~\ref{Tc-Nonmonotonic} as a function of of $\bar{\eta}' = \eta'/\kappa$. As is evident, for the case of a non-local mediator, $T_c$ behaves non-monotonically with dissipation and rises up to $40\%$ of the initial $\eta=0$ value. This happens because weakly dissipative bosons at different energy scales act coherently to give rise to an increase in $T_c$ but eventually destroy superconductivity for large dissipation. The critical value occurs when the dissipation parameter is of the order of the stiffness constant ($2\eta\sim \kappa$); at this point, $T_c$ acquires a maximum with respect to $\eta$. This physics follows from the energy integral leading to Eq.~\ref{Non-Local-GapEqn} above. To see this, notice that the role of the stiffness parameter $\kappa$ is to induce non-monotonicity in an `effective' coupling constant as a function of $\eta$ -- while $\eta$ acts only to reduce the effective coupling constant for the local case, the energy integral (leading to Eq.~\ref{Non-Local-GapEqn}) for the non-local mediator forces the gap equation to acquire dissipative contributions that both increase and decrease the effective coupling constant. Consequently, this translates into a non-monotonic behavior in $T_c$. \par
\begin{figure}[h!]
\includegraphics[width=1.7in,height=1.25in]{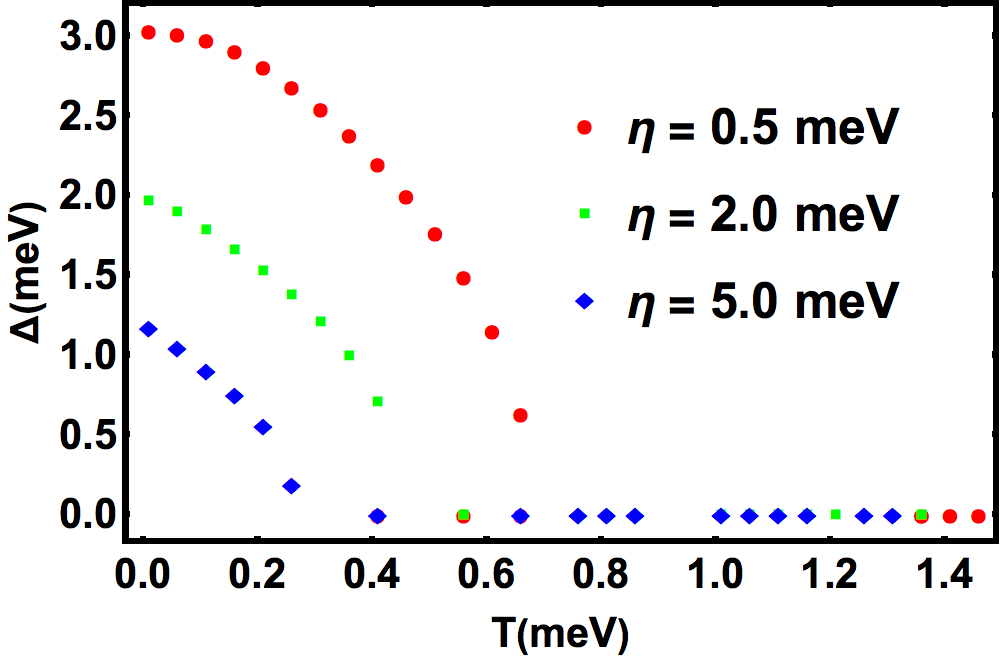}\hfill%
\includegraphics[width=1.7in,height=1.25in]{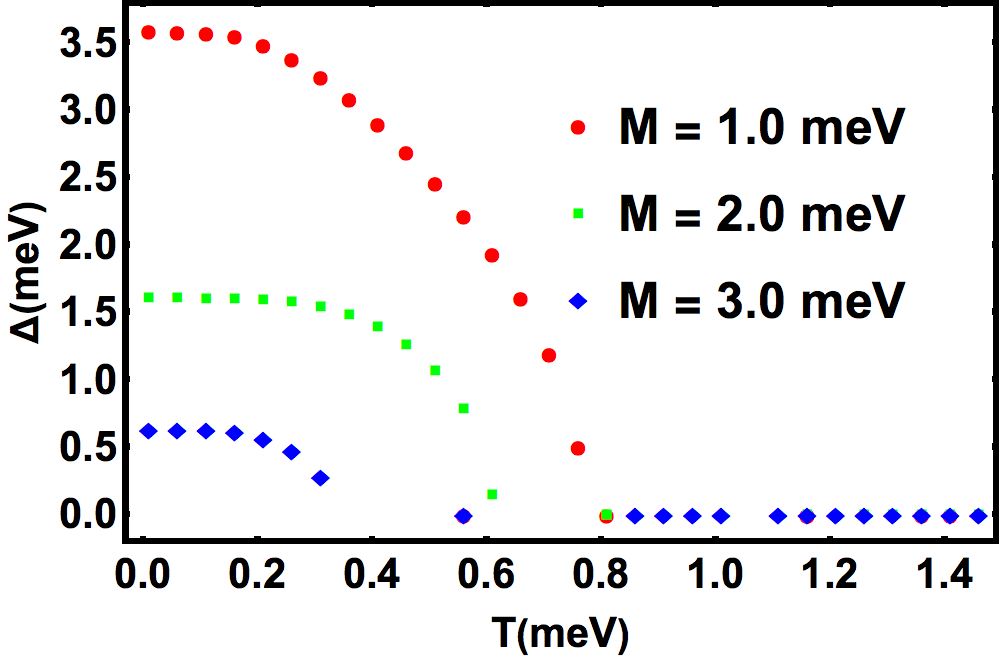}
\caption{Temperature dependence of the superconducting gap for $\lambda = 5$ meV. (Left) as a function of the dissipation parameter $\eta$ and $M=1$ meV. (Right) As a function of the mass parameter $M$ and $\eta = 0.1$ meV.  The BCS ratio $\frac{\Delta(0)}{T_c}$ increases (decreases) with the dissipation (mass) parameter.
} \label{GapVariation}
\end{figure}
\section{Gap and specific heat jump} We now study the variation of the gap with temperature and the specific heat jump at $T_c$. In Fig~\ref{GapVariation} we plot the temperature dependence of the superconducting gap as a function of the dissipation and mass parameters for $\kappa=0$. Both $\eta$ and $M$ reduce the zero temperature gap $\Delta(0)$ and $T_c$; however, dissipation (mass) has a greater (smaller) effect on $T_c$ compared to $\Delta(0)$. Hence, the BCS ratio $\frac{\Delta(0)}{T_c}$ increases (decreases) with the dissipation (mass) parameter. 
To get an analytical handle for the gap near $T_c$, we begin with the case of $\eta = M =0$ where the gap equation becomes 
\beq
1= \lambda \int_{-\infty}^{\infty} d\xi \left( \frac{1}{4 T (\xi^2 + \Delta^2)} - \frac{tanh \frac{\sqrt{\xi^2 + \Delta^2}}{2T}}{2(\xi^2 + \Delta^2)^{3/2}}\right).
\eeq
We have made use of the summation identity 
\beq
\sum_m \frac{1}{(|\omega_m|^2 + x^2) |\omega_m|^2} = \frac{x - 2 T~tanh \frac{x}{2 T}}{4 x^3 T^2}
\eeq
 above. We next expand for small gaps near $T_c$ to obtain 
\beq\nonumber
\frac{1}{\lambda} &=& \int_{-\infty}^{\infty} d\xi \Bigg[ \frac{\beta \xi - 2 tanh\left(\frac{\beta \xi}{2}\right)}{4\xi^3} \\ \nonumber
&& \frac{-3 \beta \xi + 6~ tanh\left( \frac{\beta \xi}{2}\right) + \beta \xi~tanh^2\left(\frac{\beta \xi}{2}\right)}{8\xi^5} \Delta^2 +..\Bigg]\\
&\simeq& \frac{\beta^2 a(T) }{4 \pi^2} - b \Delta^2,
\label{LocalGapVsT}
\eeq
where
\beq
a(T) = \frac{1}{2}\left[\psi\left( 2, \frac{3}{2} + \frac{\beta \Lambda}{2 \pi}\right) - \frac{1}{2}\psi(2, \frac{1}{2})\right]
\eeq
 is weakly temperature dependent in the limit of $\beta \Lambda \rightarrow \infty$, $b\simeq \frac{31}{32} \frac{\beta_c^4 \xi(5)}{\pi^4}$ and $\psi(n,x)$ is the $n$-th order digamma function. Setting the gap to zero in Eq~\ref{LocalGapVsT}, we can read off the dependence of $T_c$ on the coupling as $T_c \sim \sqrt{\lambda}$, which grows faster than the conventional BCS relation. The temperature dependence of the gap can be derived as $\Delta^2(T) = \frac{2 a(0) T_c \pi^4}{4 \pi^2 (31/32) \xi(5)} \left( T_c -T\right)$, and therefore, the normalized specific heat jump at $T_c$ is ($\gamma = 2 \pi^2 N(0)/3 $ is the normal state specific heat) $\frac{\Delta C}{\gamma T_c} = \frac{3 a(0)}{4 \xi(5)} \left(\frac{32}{31}\right) \simeq 6$, which is greater than the BCS value. Similarly, in the limit where the dissipation is much larger than the temperature and mass ($\eta |\omega_m| \gg |\omega_m|^2, M^2$), we have $\frac{1}{\lambda} \simeq \frac{u(T)}{4\pi T \eta} - \frac{v}{8 \eta \pi^4 T_c^3 }~\Delta^2$.
Here
\beq 
u(T) &=& \pi^2 - 2 \psi\left( 1, \frac{3}{2} + \frac{\beta \Lambda}{2\pi} \right), \\ \nonumber
v &=& \int_0^{\infty}  dx \Bigg[ \frac{ H(-\frac{1}{2} - i x) + c.c + Log 16}{x^4} \\ 
&&+\frac{\frac{i x}{2} \left( \psi(1, \frac{1}{2} - i x) - c.c \right)}{x^4} \Bigg],
\eeq
and $H(z)$ is the Harmonic number.
$T_c$ can be evaluated again by setting $\Delta =0$ and we see that, in this limit, $T_c \sim \frac{\lambda}{\eta}$. The cross-over from $T_c \sim \sqrt{\lambda}$ to $T_c \sim \lambda$ as a function of $\eta$ is shown in Fig~\ref{TcVsMandLambda}(right). The temperature dependence of the gap can be evaluated from above as $\Delta(T)^2 =\frac{2 \pi^5 T_c}{v}(T_c - T)$; hence, the specific heat jump at $T_c$ takes the value $\frac{\Delta C}{\gamma T_c} = \frac{3\pi^3}{v} \sim 3.64$ which is again greater than the BCS value. \par
On the other hand, expanding the gap equation for a non-local mediator ($\kappa\neq0$) in the limit of $\eta = M \rightarrow 0$, we obtain 
 $\frac{\kappa^2}{\lambda} \simeq F\left(\frac{\kappa}{2 \pi T}\right) - G\left(\frac{\kappa}{2 \pi T} \right) \tilde{\Delta}^2$. The dimensionless functions $F(x)$, $G(x)$ and $\tilde{\Delta}$ are defined as
\beq
F(x) &=& \frac{1}{2}\left[H\left( -\frac{1}{2} - i x \right) + c.c + Log~16\right], \\ \nonumber
G(x) &=& \frac{-1}{x^2}\Bigg[10 (\gamma_E + Log 4) + 5 \psi\left(0, \frac{1}{2} - i x\right) + c.c \\
&&+ i x~ \psi\left(1, \frac{1}{2} - i x\right) + c.c  - 42 x^2 \xi(3) \Bigg],
\eeq
$\tilde{\Delta} = \frac{\Delta}{2\pi T}$, and $\gamma_E$ is the Euler gamma constant. In the limit $x\ll1$, the functions $F(x)$ and $G(x)$ satisfy the property $F(x) = C_1 x^2$ and $G(x) = C_2 x^2$, where $C_1$ and $C_2$ are numerical constants. The dependence of $T_c$ on $\lambda$ goes as $T_c \sim \sqrt{\lambda}$ and the temperature dependence of the gap takes the form $\Delta(T)^2 = \frac{8\pi^2 C_1 T_c}{C_2}(T_c - T)$. This implies that the specific heat jump is $\frac{\Delta C}{\gamma T_c} \sim 5.6$, again larger than the BCS value. However, in the limit $\eta |\omega_m| \gg |\omega_m|^2, M^2 $ the expansion of the gap equation gives
\begin{align}
\frac{1}{\bar{\lambda}} = \frac{1}{2 \pi T} \Bigg[ \frac{\bar{\eta}' \pi^2}{1+ \bar{\eta}'^2 } - \frac{\bar{\eta}' \pi^4(3+ \bar{\eta}'^2)}{12(1+ \bar{\eta}'^2)} \left(\frac{\Delta}{2\pi T_c} \right)^2+ ... \Bigg]
\end{align}
where $\bar{\lambda} = \lambda/\kappa$. Setting $\Delta=0$, we see that $T_c \sim \bar{\lambda}$ and the temperature dependence of the gap is given by $\Delta(T)^2 = \frac{24T_c (T_c- T)(1+ \bar{\eta}'^2) }{(3+ \bar{\eta}'^2)}$. Hence, the specific heat jump is (weakly) dependent on the dissipation parameter and is given by $\frac{\Delta C}{\gamma T_c}  = \frac{36 (1+ \bar{\eta}'^2)}{\pi^2 (3+ \bar{\eta}'^2)}$. For small $\bar{\eta}'$, the normalized specific heat jump is $\simeq 1.2$ and is \textit{smaller} than the BCS value consistent with specific heat experiments in under-doped cuprates~\cite{Loram2001} and the pnictides~\cite{Boeri2012}. \par
\section{Discussions and experiments} Several theoretical works have explored mechanisms that yield an enhancement of $T_c$ with disorder strength. These phenomena range from competition of superconductivity with a proximate density wave phase~\cite{Nass1982, Psaltakis1984, Chubukov2012-Enhancement, Hirschfeld2016}, multiorbital effects~\cite{Andersen2017-Enhancing}, local inhomogeneities in the pairing interactions and mediators ~\cite{Andersen2018-Raising, Kivelson2005, Kivelson2003, Carlson2007, Scalettar2006, Scalettar2007} to localization~\cite{Yuzbashyan2007, Mirlin2012, Mirlin2015, Garcia-Garcia2015}. A few works have also explored the interplay between glassy phases and superconducting $T_c$~\cite{Seki1995, Larkin2002}. In~\cite{Larkin2002}, the authors study a spin-glass formed by RKKY interactions between paramagnetic spins in a superconductor, and find an interaction driven enhancement of $T_c$ for a fixed impurity density. As a function of impurity concentration, however, the authors find that the $T_c$ decreases monotonically. Ref.~\cite{Seki1995} also finds a similar decrease in $T_c$ due to a reduction of the effective interaction induced by a SG phase that does not take into account the role of dissipation explicitly.  Our results can alternatively viewed from the perspective of the well studied spin-fermion model~\cite{Pines1991, Pines1992, Stojkovic1999, Chubukov2000, Schmalian2003} where the dissipation parameter is proportional to the inverse spin-fluctuation frequency $\omega_{SF}$. In all of these works, $\omega_{SF}$ and the correlation length $\xi$ parameters are held fixed for different materials (see Table I of Ref.~\cite{Pines1991}) at $T = T_c$. Although these works do not study the effect of $\omega_{SF}$ and $\xi$ on $T_c$, legitimate questions can be raised with regards to whether these quantities can be varied by an experimentally controlled tuning knob. In the case relevant to the present context, the effect of disorder on $\omega_{SF}$ and $\xi$ needs further examination, and perhaps the current works brings forth the need for a microscopic understanding of the relationship between disorder strength and the dissipation parameters $\eta$
and  $\omega_{SF}$  . In the following paragraphs we argue for the applicability of the mechanism presented in this paper to the cuprates.  \par
We begin by emphasizing that the change in $T_c$ in our work is due to modification of the `effective' coupling by dissipation, and is unrelated to pair-breaking effects originating from lowering  translational symmetry (say due to magnetic/non-magnetic inhomogeneities, like those summarized in Ref.~\cite{Zhu2006}). Hence, it can be intuited that qualitative aspects of our conclusions must hold for higher angular momentum pairing as well (albeit with tedious calculations). This can be more readily seen by noting that when the summand in the gap equation (in Eq~\ref{GapEquation}) is decomposed into its partial fractions, there is always at least one (non-zero) term present where the dissipation parameter contributes to enhance the effective coupling strength (similar to Eq~ \ref{Non-Local-GapEqn}). This term(s) generally competes with other terms which suppress $T_c$, but gives rise to a $T_c$ increase when the dissipation is weak enough. Furthermore, according to our proposal, disorder acts as an external tuning knob of the parameter $\eta$; hence, increased irradiation leads to larger dissipation. Recent magnetization and tunnel diode (penetration depth) experiments~\cite{Welp2018} on proton irradiated LBCO at $\frac{1}{8}$ doping found up to a 50\% increase in $T_c$ as a function of radiation dosage. An increased dosage above a critical value gradually suppressed $T_c$ until the eventual destruction of superconductivity. LBCO also hosts a rich phase diagram with evidence of density wave orders (CDW, SDW~\cite{Casa2008, Tranquada2011, Dean2017} as well as spin- glass behavior~\cite{Ferretti2001} in conjunction with superconductivity in the under-doped regime. Hence, it is natural to anticipate an influence of these phases on superconductivity and examine their implications to $T_c$ variation as a function of disorder. Of the aforementioned existing mechanisms of $T_c$ enhancement proposed in literature, a competition-based scenario between superconductivity and a density wave order seems the most promising at first sight -- especially given the close proximity of the CDW phase to the superconducting dome. Indeed, this was the point of view first suggested by Leroux and co-workers in~\cite{Welp2018}. However, a closer examination of the data points to detaills that render this mechanism debatable. First, assuming that x-ray scattering is primarily sensitive to long-range CDW order~\cite{FootNote}, the CDW transition temperature seems unaffected by irradiation~\cite{Welp2018}. But a mechanism involving the competition between two mean field phases necessary involves a tug-of-war scenario where one phase gains stability at the expense of its competitor~\cite{Chubukov2012-Enhancement}. Second, it is unclear how non-magnetic disorder affects two different mean field phases (CDW, SDW, SC etc) asymmetrically in a parameter independent manner, except under very specific circumstances~\cite{Chubukov2012-Enhancement, Hirschfeld2016} which do not necessarily hold in the case of LBCO and cuprates. Third, other non-magnetic impurities are well known to kill $d$-wave superconductivity monotonically~\cite{Zhu2006}. Thus a consistent picture which distinguishes proton and electron irradiation with other point impurities like $Zn$ at a microscopic level is absent. Finally, from Anderson's theorem, one can expect that a $T_c$ enhancement that occurs through a competition based scenario must be more prevalent in $s$-wave SCs rather than higher angular momentum SCs which are far less robust to non-magnetic impurities. Experiments on the $s$-wave superconductor 2H-NbSe$_2$, however, draw conclusions that are mixed at best~\cite{Mutka1983, Crowder1975, Prozorov2018}. Hence, a reasonable explanation for non-monotonic $T_c$ dependence as a function of disorder in LBCO must necessarily involve a mechanism that \textit{does not depend on the competition of two mean-field like phases}. The proposed mechanism in this paper, along with the experimental evidence provided in the introduction, forms a feasible alternative that fits experiments.  \par
In conclusion, motivated by the close proximity of glassy phases to the superconducting dome in high $T_c$ SCs, we explored the role of dissipation on superconducting properties such as $T_c$, the temperature dependence of the gap, BCS ratio and the specific heat jump at $T_c$. We found that when the mediator is local, dissipation acts to reduce the effective coupling constant and $T_c$ monotonically. On the other hand, when the mediator is non-local, two competing effects of dissipation determine the $T_c$ variation -- first, the dissipative contributions of individual bosons at a given energy that act to suppress $T_c$, and second, collective contributions where dissipation acts to connect bosons at different energy scales that combine coherently to increase the effective coupling and $T_c$. The former (latter) contribution dominates when the dissipation parameter is greater (lesser) than the bosonic spatial stiffness, i.e., $\eta>\kappa$ ($\eta<\kappa$); $T_c$ peaks when these two scales are comparable to each other. We also studied the effects of a dissipative mediator on the  ratio $\frac{2\Delta(0)}{T_c}$ and the heat capacity jump at $T_c$, and found departures from values predicted by BCS theory. In particular, the specific heat jump at $T_c$ acquires a value smaller than that predicted by BCS theory when the mediator is both dissipative and non-local, consistent with experiment. We pointed out consequences of our results to recent proton irradiation experiments in LBCO~\cite{Welp2018} where superconducting $T_c$ is enhanced with increased radiation disorder despite a robust CDW transition temperature, and concluded that one does not require a `tug-of-war' like scenario between two competing phases to enhance superconductivity.   \par
\textit{Acknowledgements:} We thank  P. J. Hirschfeld and P. W. Phillips for discussions. This work is supported by the DOE grant number DE-FG02-05ER46236.

 \bibliographystyle{apsrev4-1}
\bibliography{Dissipation.bib}

\begin{thebibliography}{64}
\bibitem [{Foo()}]{FootNote}%
  \BibitemOpen
  \href@noop {} {\bibinfo  {journal} {It is possible that short-range CDW
  correlations are picked up by x-rays}\ }\BibitemShut {NoStop}%
\end{thebibliography}
\end{document}